\documentclass[preprint2]{aastex} 
\usepackage{graphicx,natbib} 
 
\shorttitle{VLBI Astrometric Search for a Companion of TVLM\,513--46546}
\shortauthors{Forbrich et al.}
 
\begin{document} 
 
\title{An Astrometric Search for a Sub-stellar Companion of the M8.5
Dwarf TVLM\,513--46546 Using Very Long Baseline Interferometry}
 
\author{
Jan Forbrich\altaffilmark{1,2}, 
Edo Berger\altaffilmark{1},
\& Mark J. Reid\altaffilmark{1}
}          
 
\altaffiltext{1}{Harvard-Smithsonian Center for Astrophysics, 60
Garden Street, Cambridge, MA 02138, USA}
\altaffiltext{2}{University of Vienna, Department of Astrophysics,
T\"urkenschanzstr. 17, 1180 Vienna, Austria}
 
\begin{abstract} We conducted multi-epoch VLBI observations to search
for astrometric reflex motion caused by a sub-stellar companion of the
M8.5 dwarf TVLM\,513--46546.  The observations yield an absolute
parallax corresponding to a distance of $10.762\pm0.027$~pc and a
proper motion of $78.09\pm0.17$~mas\,yr$^{-1}$.  From the absence of
significant residual motion, we place an upper limit to any reflex
motion caused by a companion, extending the parameter space covered by
previous near-infrared direct-imaging searches.  By covering different
orbital periods, the data exclude a phase-space of companion masses
and orbital periods ranging from 3.8~$M_{\rm jup}$ with an orbital
radius of $\sim0.05$~AU (orbital period of 16 days) to 0.3~$M_{\rm
jup}$ with an orbital radius of $\sim0.7$~AU (orbital period of 710
days). \end{abstract}
 

\section{Introduction} 
\label{sec:intro}

Searches for extrasolar planets over the past decade have successfully
utilized a wide range of techniques, including radial velocity,
transits, gravitational microlensing, direct imaging, and even pulsar
timing \citep{wol92,may95,cha00,bon04,kal08}.  The various techniques provide 
sensitivity in
different ranges of the orbital period and exoplanet mass phase-space,
as well as in terms of the primary star's brightness.  In
particular, radial velocity and transit searches of very low mass
stars and brown dwarfs are quite challenging due to the faintness of
these objects, their broad (molecular) spectral features, and their
ubiquitous variability; indeed only a few exoplanet host stars
with masses of $\lesssim 0.2$ M$_\odot$ are currently known
\citep{cbi+09,kbb+12,mja+12}.  An additional technique, which has not yielded
exoplanet detections so far, is optical/infrared astrometry (\citealt{psl05};
although see the recent detection of a $28$ M$_{\rm jup}$ companion to an L1.5 
dwarf: \citealt{sls+13}) relying on the positional shift of the star around 
the center of mass of the orbit (reflex motion).  This technique in principle
has the advantage that it is insensitive to the inclination of the exoplanet
orbit.

Astrometric searches with the requisite sensitivity to detect
exoplanets can also be carried out at radio frequencies using very
long baseline interferometry (VLBI); see \citet{bbf+09,fob09}.  In
this context, the discovery of radio emission from late-M and L dwarfs
\citep{bbb+01,ber06} provides a unique opportunity to uncover
exoplanets or brown dwarf companions to stars and brown dwarfs with
$M\lesssim 0.1$ M$_\odot$.  Our recent detection of the M8.5 dwarf
TVLM\,513--46546 (hereafter, TVLM\,513) with VLBI \citep{fob09},
motivated us to perform an astrometric search for a sub-stellar
companion using long-term VLBI observations.  A previous direct
imaging search for a companion to TVLM\,513 \citep{clo03} yielded no
detections at separations of $0.1-15''$, corresponding to about
$1-160$ AU at the distance of the source.

In this paper we report the results of our multi-epoch VLBI program,
carried out with the NRAO Very Long Baseline Array (VLBA).  The
observing sequence was designed to sample a range of timescales
appropriate for measurements of parallax, proper motion, and any
reflex motion from a sub-stellar companion.  The observations thus
allow us to both significantly improve on the optical measurements of
the parallax and proper motion, and to detect a sub-stellar companion
over a wide range of orbital separations in the sub-AU range.  We
summarize the known properties of TVLM\,513 in \S\ref{sec:tvlm}, and
describe the VLBI observations and analysis in \S\ref{sec:obs}.  Our
measurements of an improved parallax and proper motion are described
in \S\ref{sec:parpm}, and the results of the astrometric companion
search are discussed in \S\ref{sec:astrom}.

\section{TVLM\,513--46546}
\label{sec:tvlm}

\begin{deluxetable}{lllrrrr} 
\tabletypesize{\scriptsize} 
\tablecaption{Observing log and detections\label{tab_obslog}} 
\tablewidth{0pt} 
\tablehead{ 
\colhead{ID BF100}  & 
\colhead{Date}      & 
\colhead{UT Time}   & 
\colhead{Flux}      & 
\colhead{rms}       & 
\colhead{SNR}       & 
\colhead{RA, Dec}   \\ 
\colhead{}          &  
\colhead{}          & 
\colhead{}          & 
\colhead{(mJy)}     & 
\colhead{(mJy)}     & 
\colhead{}          & 
\colhead{(J2000)}                                                             
} 
\startdata 
A & 18-MAR-2010 & 07:58--13:00     & 	 0.324 & 0.046 &  7.1 & 15 01 08.1596568$\pm$32  +22 50 01.49944$\pm$12\\  
B & 26-MAR-2010 & 07:26--12:28     & 	 0.286 & 0.045 &  6.4 & 15 01 08.1589695$\pm$47  +22 50 01.50580$\pm$12\\  
C & 05-APR-2010 & 06:46--11:49     & 	 0.894 & 0.046 & 19.7 & 15 01 08.1580103$\pm$15  +22 50 01.51294$\pm$04\\  
I & 26-APR-2010 & 04:59--12:00     & 	 0.274 & 0.033 &  8.4 & 15 01 08.1556973$\pm$29  +22 50 01.52403$\pm$08\\  
D & 27-MAY-2010 & 03:22--08:25     & 	 0.349 & 0.052 &  6.8 & 15 01 08.1520411$\pm$41  +22 50 01.52852$\pm$15\\  
E & 25-JUN-2010 & 01:28--06:31     &  $<$0.162 & 0.054 & (3$\sigma$) & 					    ---\\ 
F & 03-NOV-2010 & 16:49--21:52     &  $<$0.157 & 0.052 & (3$\sigma$) & 					    ---\\  
G & 08-MAR-2011 & 08:39--13:43     & 	 0.155 & 0.045 &  3.4 & 15 01 08.1572406$\pm$121 +22 50 01.42538$\pm$32\\  
H & 03-AUG-2011 & 22:53--03:57(+1) &  $<$0.163 & 0.054 & (3$\sigma$) & 					    ---\\  
\enddata 
\end{deluxetable} 
 
\begin{deluxetable}{lr} 
\tablecaption{Parallax fit results\label{tab_parfit}} 
\tablewidth{0pt} 
\tablehead{ 
\colhead{Parameter}    & 
\colhead{BB251+ABCIDG} \\
} 
\startdata 
$\pi_{\rm abs}$ (mas)                & $ 92.92\pm0.23$\\
$\mu_{\rm RA,abs}$ (mas~yr$^{-1}$)   & $-42.56\pm0.12$\\
$\mu_{\rm Dec,abs}$ (mas~yr$^{-1}$)  & $-65.47\pm0.12$\\
$\mu_{\rm abs}$ (mas~yr$^{-1}$)      & $ 78.09\pm0.17$\\
$\phi$ (deg)                         & $213.03\pm0.12$\\
$d$ (pc)                             & $10.762\pm0.027$\\
$v$ (km s$^{-1}$)                    & $3.987\pm0.010$\\
$R(\pi_{\rm abs})$                   & $0.695$\\ 
$R(\mu_{\rm RA,abs})$                & $0.007$\\ 
$R(\mu_{\rm Dec,abs})$               & $0.015$\\ 
\enddata 
\end{deluxetable}

An optical parallax of TVLM\,513, an M8.5 star \citep{tin93}, was
first obtained by \citet{tin95} who derive a relative parallax of
$\pi_{\rm rel} = 100.5\pm5.2$~mas, as well as a relative proper motion
of $\mu_{\rm rel}=85.0\pm7.8$~mas toward position angle $\phi_{\rm
rel} =205.8\pm 4\fdg 2$ (East of North). From an estimate of the
photometric distances to the reference stars, they derive an absolute
parallax of $\pi_{\rm abs} = 101.8\pm5.2$~mas. Subsequently,
\citet{dah02} improved this result with a similar technique, obtaining
$\pi_{\rm rel} = 93.2\pm0.5$~mas and a proper motion of $\mu_{\rm rel}
=62.9\pm0.4$~mas with a position angle of $\phi_{\rm rel} =203.0\pm
0\fdg 3$. Their derived absolute parallax is $\pi_{\rm abs} = 94.4\pm
0.6$~mas, and the tangential velocity of the star is $v_{\rm tan}=
3.2$~km\,s$^{-1}$.

TVLM\,513 has also been detected on several occassions at radio
frequencies, and is known to be variable at 8.4 GHz with an average
flux density of about $300$ $\mu$Jy
\citep{ber02,ost06,hal06,hal07,ber08,fob09,doy10}.

\section{Observations and Calibration} 
\label{sec:obs}

Following our initial detection of TVLM\,513 in VLBI observations
\citep{fob09}, we observed the source at nine more epochs in 2010 and
2011 using the NRAO Very Long Baseline Array (project BF100 with
tracks A-I).  At one of these epochs (I), we added the 100-m NRAO
Green Bank Telescope (GBT) to increase the sensitivity. Except for
epoch I, which had a duration of 7 hours, the observations lasted for
5 hours each. The observing log is listed in Table~\ref{tab_obslog}.
We used the initial observations to estimate the offset of TVLM\,513
(with respect to the background sources), and the optical parallax and
proper motion of \citet{dah02} to predict positions for the
interferometer correlations for subsequent observations. The observing
sequence was designed to provide sensitivity to different masses and
orbital periods for a potential sub-stellar companion.

We calibrated the raw interferometer visibility data with the AIPS
software package using standard procedures. To improve the astrometric
accuracy, we conducted ``geodetic block'' observations at 8.4 GHz of
ICRF sources with well-determined positions \citep{rei04}. These
blocks were placed at the beginning and end of each observing track.
For epoch I, which included the Green Bank Telescope, we inserted an
additional block near the middle of the track.  Ionospheric delays
were removed using Global Positioning System total electron content
measurements and the data were corrected using the best Earth
orientation parameter values.  Standard procedures included removing
residual delays as clock drifts and zenith atmospheric delays. We
removed electronic delays and differences between the IF bands using
observations of a strong calibration source, J1800+782.  Subsequently,
we interpolated the phases for the rapid-switching reference source
J1455+2131 (located $1.84^\circ$ away from the target) to cover the
times of TVLM\,513 scans and removed them from the latter.

\begin{figure*}
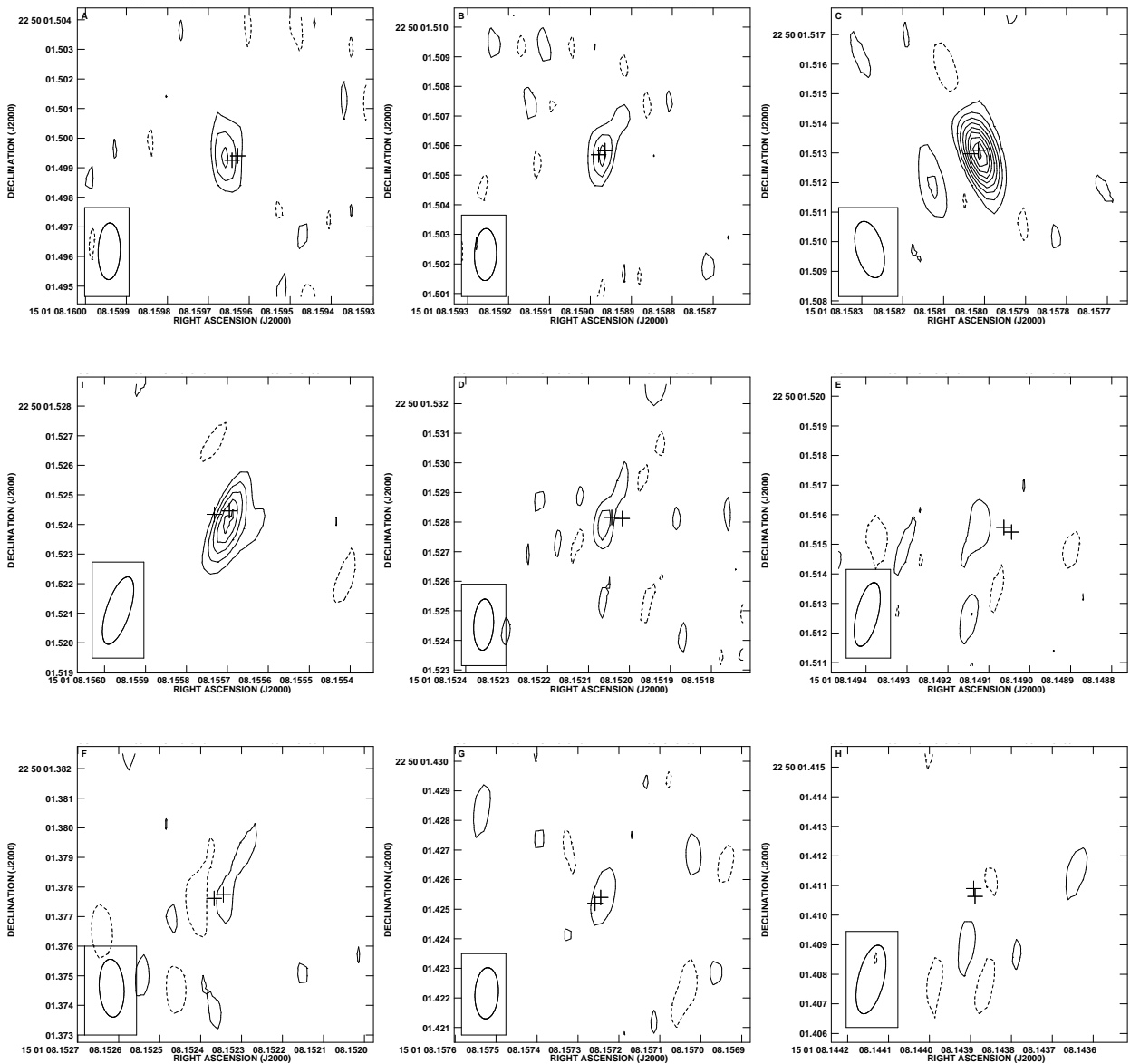

\begin{minipage}{0.33\linewidth} 
\includegraphics*[angle=-90,bb=528 132 69 666,width=1.0\linewidth]{DWF1501AX19.PS} 
\end{minipage} 
\begin{minipage}{0.33\linewidth} 
\includegraphics*[angle=-90,bb=528 132 69 666,width=1.0\linewidth]{DWF1501BX19.PS} 
\end{minipage}
\begin{minipage}{0.33\linewidth} 
\includegraphics*[angle=-90,bb=528 132 69 666,width=1.0\linewidth]{DWF1501CX19.PS} 
\end{minipage}
 
\vspace*{5cm}
\begin{minipage}{0.33\linewidth} 
\includegraphics*[angle=-90,bb=528 132 69 666,width=1.0\linewidth]{DWF1501IX19.PS} 
\end{minipage} 
\begin{minipage}{0.33\linewidth} 
\includegraphics*[angle=-90,bb=528 132 69 666,width=1.0\linewidth]{DWF1501DX19.PS} 
\end{minipage} 
\begin{minipage}{0.33\linewidth} 
\includegraphics*[angle=-90,bb=528 132 69 666,width=1.0\linewidth]{DWF1501EX19.PS} 
\end{minipage}

\vspace*{5cm}
\begin{minipage}{0.33\linewidth} 
\includegraphics*[angle=-90,bb=528 132 69 666,width=1.0\linewidth]{DWF1501FX19.PS} 
\end{minipage} 
\begin{minipage}{0.33\linewidth} 
\includegraphics*[angle=-90,bb=528 132 69 666,width=1.0\linewidth]{DWF1501GX19.PS} 
\end{minipage} 
\begin{minipage}{0.33\linewidth} 
\includegraphics*[angle=-90,bb=528 132 69 666,width=1.0\linewidth]{DWF1501HX19.PS} 
\end{minipage} 
\caption{Contour plots of TVLM\,513 in epochs ABCIDEFGH, starting in
the upper left. The contour spacing is in factors of 2, starting at
$\pm2\sigma$. The `+' symbols mark the expected position, according to
the model fit BB251+ABCID, at the beginning and at the end of the
observation. With the exception of epoch F, the expected motion is
westward. The synthesized beam size is shown in the lower left corner
of each panel. \label{fig_kntr}} 
\end{figure*}

We imaged the calibrated data with the AIPS task IMAGR, with a typical
synthesized beam FWHM of 2.0$\times$0.8~mas (for ROBUST=0 weighting),
elongated in roughly the North-South direction, depending on the
scheduling of each epoch.  Finally, we estimated the source positions
by fitting 2-dimensional Gaussian brightness distributions in the
image plane.  Maps of TVLM\,513 for all epochs, with depictions of the
synthesized beam sizes, are shown in Fig.~\ref{fig_kntr}. The results
of the fits to determine the peak flux density and the source
positions are listed in Table~\ref{tab_obslog}.

We note that the parallax and proper motion of the source are large
enough to cause an apparent movement of up to 0.07~mas\,hr$^{-1}$,
i.e., 9\% of the typical minor axis of the synthesized beam.  Thus,
the precise times of day of the observations were taken into account.
 
\clearpage

\clearpage
 
\begin{figure} 
\includegraphics[angle=0,width=1.0\linewidth,bb=83 50 265 223]{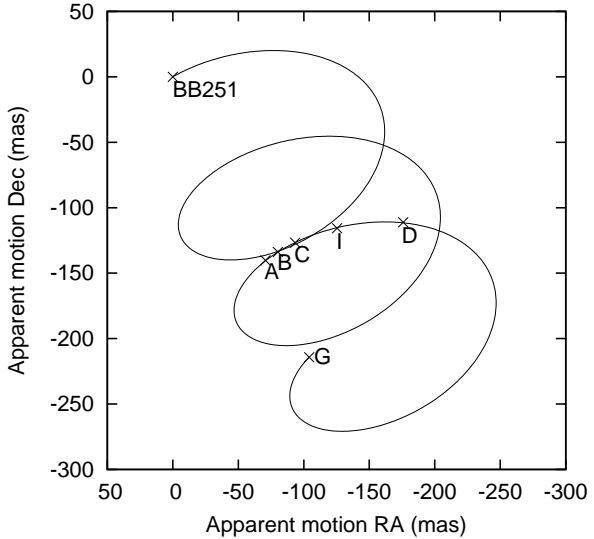}
\caption{Parallax and proper motion fit to all seven epochs with
detections of TVLM\,513, including the observation reported by
\citet{fob09}, marked as crosses. The line shows the model. The
positional uncertainties are too small to be seen on this
scale. 
\label{fig_model}}
\end{figure} 
 
\begin{figure} 
\includegraphics[angle=0,width=1.0\linewidth,bb=83 50 265 223]{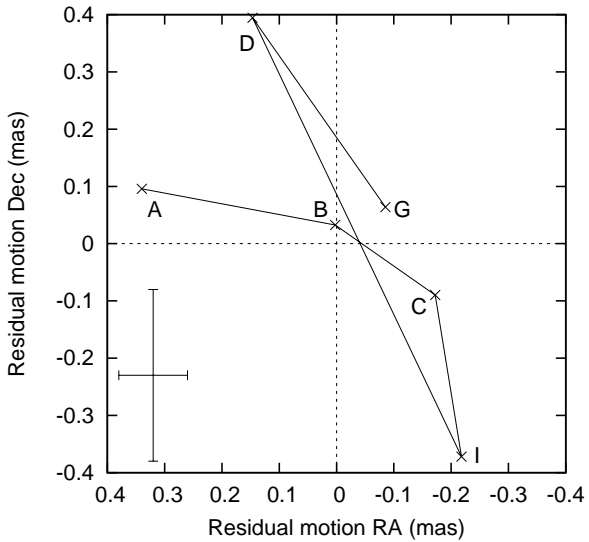}
\caption{Residuals of the parallax and proper motion fit for the
BB251+ABCIDG fit, shown for the 2010 observations. The typical thermal
position uncertainty is shown in the lower left
corner. 
\label{fig_results2a}} 
\end{figure}
 
\begin{figure} 
\includegraphics[angle=0,width=1.0\linewidth,bb=83 50 265 223]{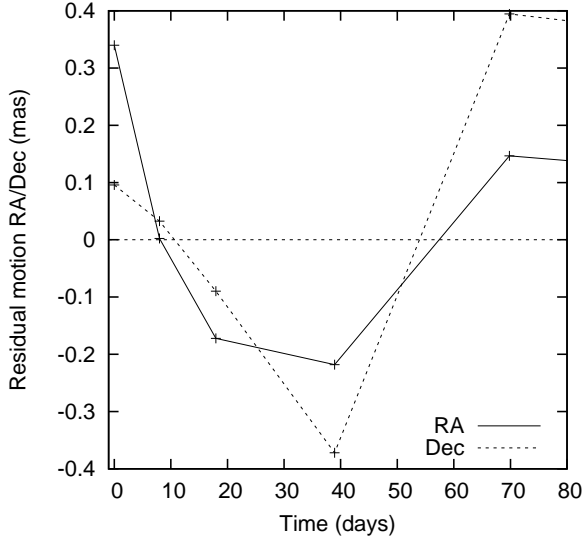}
\caption{Same as Figure~\ref{fig_results2a} but with the RA (solid)
and Dec (dashed) residuals plotted as a function of observing
time. Residuals of the parallax and proper motion fit for the 2010
observations (ABCIDG). The typical errors are the same as in
Figure~\ref{fig_results2a}. 
\label{fig_results2b}}
\end{figure} 
 
\begin{figure} 
\includegraphics[angle=0, width=1.0\linewidth, bb=50 50 256 199]{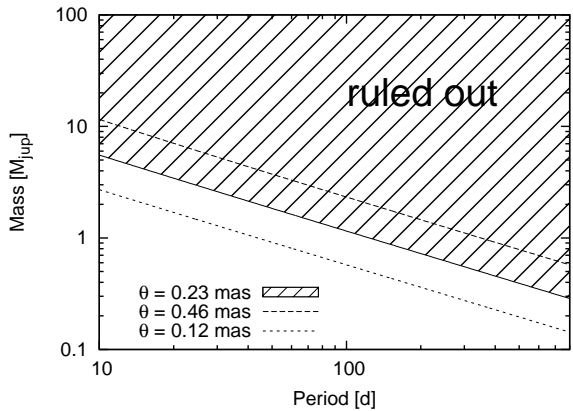}
\caption{Plot of the minimum mass of a single companion that can be
ruled out (hashed area), as a function of orbital period, as
calculated from Equation~\ref{eqn:theta}. The solid line shows the
limit for the error floor of 0.23~mas while the two dashed lines show
the corresponding limit for error floors higher and lower by a factor
of 2, to indicate sensitivity. The range covered corresponds to
orbital radii ranging from $\sim0.05$~AU to $\sim0.7$~AU.
\label{fig_results3}}
\end{figure}

We detect TVLM\,513 with a signal-to-noise ratio (SNR) $>5$ only in
the first five epochs (ABCID), collected within a two month span.
Based on a parallax and proper motion fit using epochs BB251
\citep{fob09} and ABCID, we searched images from the subsequent epochs
EFGH for lower-significance detections at the predicted positions.  We
found that the source was detected in epoch G with ${\rm SNR}=3.4$.
In epoch E, weak emission is seen offset by two beams from the
predicted position (see Fig.~\ref{fig_kntr}), too far off to be
considered a real detection of TVLM\,513.

TVLM\,513 is clearly variable with the brightest detection in epoch C
at almost 0.9~mJy (19.7$\sigma$). Among the first five epochs, TVLM
513 is variable by a factor of $\sim$3.4. In fact, the quiescent flux
density reported by \citet{ost06}, at 228$\pm$11~$\mu$Jy, is only
slightly above our 3$\sigma$ upper limits. The 3.4$\sigma$ detection
in epoch G is at 155$\pm$45~$\mu$Jy.

\section{Parallax and Proper Motion} 
\label{sec:parpm}
 
Combining all detections, we estimate the absolute parallax and proper
motion of TVLM\,513.  The fit results for the parallax and the
absolute proper motion are listed in Table~\ref{tab_parfit}.  Also
listed are the multiple correlation coefficients $R$ of these
parameters which are taken into account in the determination of the
parameter errors.

To account for systematic errors, from residual atmospheric phase
effects as well as ummodeled motion from an unseen companion, we
introduce error floors for the positional uncertainties such that
post-fit residuals yielded a reduced $\chi_{\rm red}^2=1$ in both
coordinates.  This resulted in error floors of $\approx0.23$ mas,
which are considerably larger than the formal positional errors from
the Gaussian fits (governed only by SNR), indicating that the total
error is dominated by unmodeled effects.
 
The measured parallax of $\pi_{\rm abs}=92.92\pm0.23$~mas corresponds
to a distance of $d=10.762\pm0.027$~pc. The total absolute proper
motion is $\mu_{\rm abs}=78.09\pm0.17$~mas toward a position angle
$\phi = 213.03\pm0\fdg 12$.  At the derived distance, the proper
motion translates into a tangential velocity of $v_{\rm tan} =
3.987\pm 0.010$~km\,s$^{-1}$.
 
Our determination of the parallax is slightly lower than the parallax
determined from optical measurements by \citet{dah02}, but the two
measurements agree within $\approx2\sigma$ joint-error.  The main
advantage of our measurement is that it is directly tied to an
absolute reference frame, via phase referencing to the compact
extragalactic source J1455+2131.

\section{Astrometric Companion Search} 
\label{sec:astrom}

The residuals of the parallax and proper motion fit are a measure of
the (systematic-dominated) measurement error and any orbital reflex
motion caused by potential sub-stellar companions.  As shown in
Figure~\ref{fig_results2a}, the residuals are smaller than 0.5~mas.
The average {\it thermal} position uncertainty \footnote{The thermal
position uncertainty is calculated as a fraction of the synthesized
beam size as 0.5 FWHM/SNR, where FWHM is the beam size and SNR is the
signal to noise ratio (e.g., \citealp{rei88}).}  of
$\sim60\times150$~mas (in the shape of the synthesized beam) is
clearly insufficient to explain the residuals, especially in the
easterly direction.

Plotted in temporal order, the residuals may indicate an orderly
motion. In Figures~\ref{fig_results2a} and \ref{fig_results2b}, the
residuals are shown as a function of time and separately for RA and
Dec, focusing on the 2010 epochs ABCID and the corresponding fit.

For the companion search, we consider the radii of circular orbits and
do not consider inclination since our interest is only in the maximum
reflex motion.  The reflex motion $\theta$, a function of the orbital
radius $a$, can be expressed as a function of the observationally
constrained orbital period $P$ using Kepler's Third Law:
\begin{equation}
\theta = \frac{a}{d} \frac{m_c}{m_*} = \frac{m_c}{d}
\left(\frac{GP^2}{4\pi^2(m_*+m_c)^2}\right)^{1/3},
\label{eqn:theta}
\end{equation}
where $d$ is the distance, $m_c$ is the companion mass, and $m_*$ is
mass of the central object. Assuming a mass of TVLM\,513 of
0.08~$M_\odot$, we can thus rule out a certain range of single
sub-stellar companions, depending on their mass and orbital radius.

The pattern of residuals might be consistent with an orbital period of
$\approx70$ days and an amplitude of $\sim0.4$~mas which would
correspond to a companion mass of 2.5~$M_{\rm jup}$ with an orbital
radius of 0.14~AU. However, conservatively, we only use the residuals
to bound ranges of companion masses and semi-major axes (or orbital
periods).

We use a total systematic error of 0.23~mas, as derived above, as an
upper limit to any reflex motion caused by an assumed single
companion.  To make a statement on which mass range of (single)
sub-stellar companions can be ruled out by our experiment, we also
need to consider the sensitivity for certain time scales, as covered
by the observations.  Since the A and D observations are separated by
70 days, and since the maximum reflex motion is caused within half the
orbital period, this corresponds to a sensitivity to periods of 140
days.  Including epoch G, this sensitivity increases to 710 days, but
the intervening time is not well sampled.  The shortest time scale is
set by the 8 days interval between observations A and B, corresponding
to an orbital period of 16 days.  Our experiment has some sensitivity
also to shorter periods since any such companion would still increase
the apparent astrometric noise in all epochs.
 
In Figure~\ref{fig_results3}, we thus show the phase space that we can
confidently rule out for single sub-stellar companions orbiting TVLM
513, as calculated from Equation~\ref{eqn:theta}.  With the time
scales covered by epochs ABCID, our upper limit rules out a companion
of 0.9~$M_{\rm jup}$ with an orbital period of 140 days and a
companion of 3.8~$M_{\rm jup}$ with an orbital period of 16 days. For
epochs ABCIDG, the mass sensitivity drops to 0.3~$M_{\rm jup}$,
underlining the power of this method.
 
Previously, \citet{clo03} have used near-infrared adaptive-optics
imaging to rule out companions at separation of $\gtrsim 0\farcs 1$,
corresponding to a projected orbital radius of $\gtrsim 1$ AU.  The
largest orbital radii that we rule out are a 3.8~$M_{\rm jup}$
companion with an orbital radius of $\sim0.05$~AU or a 0.3~$M_{\rm
jup}$ companion with an orbital radius of $\sim0.7$~AU when including
epoch G. All of these orbits are smaller than the orbits that could be
resolved by direct imaging observations.

\section{Conclusions} 
 
A multi-epoch phase-referenced VLBI observation of TVLM\,513 yields
detections in six out of nine epochs. The observations were spaced in
different intervals to search for residual reflex motion due to an
unseen companion, with sensitivity to a wide range of orbits.  A fit
of the apparent motion to solve for the parallax and absolute proper
motion yields a distance of $d=10.762\pm0.027$~pc with an accuracy of
0.25\%. After accounting for the parallax and proper motion of TVLM
513, any residuals of this fit could be caused by unseen sub-stellar
companions. We find a suggestive pattern in the residual motion which
is close to the noise limit, and we use these residual as a
measurement of the total systematic error, including any reflex
motion. In this way, we obtain an upper limit to the reflex motion as
a function of orbital period that would be caused by companions with
masses higher than $3.8$~M$_{\rm jup}$ or $0.3$~M$_{\rm jup}$ with
corresponding orbital periods of 16 and 710 days, respectively.  The
mass sensitivity could be improved to lower-mass companions with
continued monitoring of this source even if in the case of TVLM\,513,
observations have been hampered by variability. More generally, the
sensitivity upgrade of the VLBA is now enabling such observations
toward a larger number of sources.

\acknowledgments{The National Radio Astronomy Observatory is a
facility of the National Science Foundation operated under cooperative
agreement by Associated Universities, Inc. E.~B.~acknowledges support for this work from the National Science Foundation through Grant AST-1008361.}
 
{\it Facilities:} \facility{VLBA}

\bibliographystyle{apj}
\bibliography{bib_bdvlbi}


\end{document}